\documentclass[aps,prb,showpacs,twocolumn,superscriptaddress]{revtex4}
\usepackage{graphicx}
\usepackage{bm}
\usepackage{amsmath}

\def\be{\begin{equation}}       \def\ee{\end{equation}}
\def\bea{\begin{eqnarray}}      \def\eea{\end{eqnarray}}

\begin{document}

\title{
Universal phase diagrams for the quantum spin Hall systems
}
\author{Shuichi Murakami}
\email[Electronic address: ]{murakami@stat.phys.titech.ac.jp}
\affiliation{Department of Physics, Tokyo Institute of Technology,
2-12-1 Ookayama, Meguro-ku, Tokyo 152-8551, Japan}
\affiliation{
PRESTO, Japan Science and Technology Agency (JST), Kawaguchi, 
Saitama, 332-0012, 
Japan}
\author{Shun-ichi Kuga}
\affiliation{Department of Applied Physics, University of Tokyo,
7-3-1 Hongo, Bunkyo-ku, Tokyo 113-8656, Japan}

\begin{abstract}
We describe how the three-dimensional 
quantum spin Hall phase arises from the insulator
phase by changing an external parameter. In 3D systems without inversion 
symmetry, a gapless phase should appear between the two phases
with a bulk gap. The gapless points are
monopoles and antimonopoles (in $\mathbf{k}$ space), whose topological
nature is the source of this gapless phase. 
In general, when the external parameter
is changed from the ordinary insulator phase, 
two monopole-antimonopole pairs are created and the system
becomes gapless. The gap-closing points (monopoles and antimonopoles) 
then move in the $\mathbf{k}$ space as the parameter
is changed further. They eventually annihilate in pairs, with
changing partners from the pair creations, and the system opens
a gap again, entering into the quantum spin Hall phase.
\end{abstract}

\maketitle

\section{Introduction}
Spin Hall effect (SHE) \cite{Murakami03a,Sinova04}
has been attracting current interest,
because it enables us to produce spin current without 
magnetic field or magnet. The key aspect of this phenomenon
is that the spin current is time-reversal invariant, unlike the 
spin itself. Due to this fact, the spin current can be 
induced without breaking the time-reversal symmetry. 
The physics of spin current opens up a new field for the spintronics.

In addition to the SHE in metals and doped semiconductors,
SHE in insulators\cite{Murakami04c}, 
including the quantum spin Hall (QSH) effect 
\cite{Kane05a,Kane05b, Bernevig05a}, has been studied intensively.
The quantum spin Hall phase in two dimensions (2D) 
is gapped in the bulk, while it 
has gapless edge modes carrying spin current without breaking time-reversal 
symmetry.
The interesting point is that these edge modes are topologically 
protected. They are robust against weak disorder or interaction 
\cite{Wu05,Xu05}.
Although edge states are usually sensitive to boundary conditions
such as surface roughness and impurities,
the present gapless edge states survive even if the boundary condition 
is changed.
This topological protection comes from topological order in 
the bulk, which is characterized by
 the $Z_2$ topological number $\nu$.
$\nu$ takes only two values two values $\nu\equiv 0$ (mod 2) ($\nu=$even) 
and $\nu\equiv 1$ 
(mod 2) ($\nu=$odd). 
$\nu\equiv 0$ and $\nu\equiv 1$ correspond to the 
ordinary insulator (I) phase and the QSH phase, respectively.  
The Z$_2$ topological number $\nu$ represents the number of Kramers pairs
of edge states. 
This phase has been proposed theoretically in bismuth thin film 
\cite{Murakami06b}. It has also been proposed theoretically  
in CdTe/HgTe/CdTe quantum
well\cite{Bernevig06f}, 
and it was demonstrated experimentally \cite{Konig07}.
Similar effect has been proposed theoretically for
three dimensions (3D) \cite{Fu06b,Moore06}, 
and is demonstrated in Bi$_{0.9}$Sb$_{0.1}$ \cite{Hsieh08}. 
The following property of the $Z_2$ topological number
is important.
When the bulk states are gapped, this topological 
number will not change as far as the interaction or nonmagnetic
disorder is not strong enough to close the bulk gap, or to 
break to time-reversal symmetry spontaneously.

We note that 
the $Z_2$ topological number is encoded in the physics of gap-closing.
The $Z_2$ topological number is defined as a Pfaffian of the matrix of 
the
time-reversal operator, which 
involves the phase of the wavefunctions over the whole Brillouin zone.
Its calculation is involved, and its physical meaning is 
hard to understood in a intuitive way.
On the other hand, if we focus on the change of the 
$Z_2$ topological number occuring at the QSH-I phase transition, 
the change involves only the local information 
in the $\mathbf{k}$ space, and is much simpler.
Thus by studying how the phase transition between
the QSH and the ordinary insulating phases occur, we
can get deeper insight into the $Z_2$ topological number. 
This transition necessarily accompanies closing of the bulk gap. 
We note that the gap closing is not
so trivial as it looks. Suppose we change one parameter in
the system and check whether the gap closes or not.
Because of the level repulsion, in many cases the gap 
does not close due to various 
matrix elements for interband hybridization. 
These matrix elements should vanish 
simultaneously, in order to close the gap. 
In some exceptional cases the gap closes; the conditions
for the exceptional cases are related with the $Z_2$ topological number, 
and these are what we pursue in this paper.

In the theory of gap closing by tuning an external parameter, 
momenta which satisfy 
$\mathbf{k}\equiv -\mathbf{k}$ (mod $\mathbf{G}$) play an important role,
where 
$\mathbf{G}$ is a reciprocal lattice vector. 
Such momenta are called 
the time-reversal invariant momenta (TRIM) $\bm{\Gamma}_i$, 
and have the values $\bm{\Gamma}_{i=(n_1n_2n_3)}=(
n_1\mathbf{b}_{1}
+n_2\mathbf{b}_{2}
+n_3\mathbf{b}_{3})/2$  in 3D, and 
$\bm{\Gamma}_{i=(n_1n_2)}=(
n_1\mathbf{b}_{1}
+n_2\mathbf{b}_{2})/2$ in 2D, where
$n_j=0,1$ and $\mathbf{b}_{j}$ are primitive reciprocal lattice vectors.  
The TRIM are the momenta with $\bm{\Gamma}_i=\mathbf{G}/2$ where 
$\mathbf{G}$ is a reciprocal lattice vector including zero vector. 
The TRIM are crucial in the sense that they are invariant 
under time reversal.
It has been revealed through the research of $Z_2$ topological numbers 
\cite{Kane05a,Fu06a} and gap-closing \cite{Murakami07a,Murakami07b}that the situation is quite different between the systems with and without 
inversion-(${\cal I}$-) symmetry 

\begin{figure}
\includegraphics[scale=0.5]{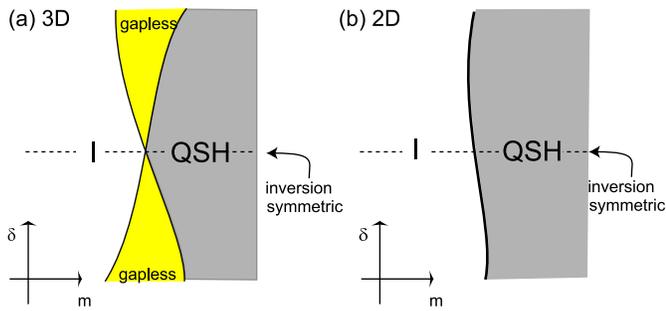}
\caption{(Color online) Phase diagram for the quantum spin Hall (QSH) and insulator (I)
phases in (a) 3D and (b) 2D. 
$m$ is a parameter driving the phase transition, and $\delta$ represents
breaking of inversion symmetry. $\delta=0$ corresponds
to the inversion-symmetric system.} 
\label{fig:phase-diagram-3D}\end{figure}

By studying the gap-closing, one can see how the phase transition 
between the QSH and insulator phases  occurs.
We studied the 2D system in Ref.~\onlinecite{Murakami07a}, and we have
obtained the universal phase diagram 
(Fig.~\ref{fig:phase-diagram-3D} (b)). 
The 3D system is studied  
in Ref.~\onlinecite{Murakami07b} and we give a handwaving argument to 
deduce the universal phase diagram in 3D, as shown in 
Fig.~\ref{fig:phase-diagram-3D} (a). 
However, general theory for the 3D systems is still lacking,
particularly for the ${\cal I}$-asymmetric systems.
In this paper we describe the phase transition between the QSH and insulator
phases in 3D and characterize its topological nature 
in a generic context. 
We describe the gap closing in 
the $\mathbf{k}$ space for ${\cal I}$-asymmetric systems.
From the topological characterization in this paper, 
we prove that the gapless phase necessarily comes in 
between the two phases. 
This paper is organized as follows. In Section \ref{sec:2} we describe how
the phase transition between the quantum spin Hall and insulator phases occurs.
Section \ref{sec:3} is devoted to a calculation on the three-dimensional 
Fu-Kane-Mele model to verify the results in the previous section.
In Section \ref{sec:4} we give conclusions and discussions.

We henceforth consider only clean systems without any impurities or
disorder, while the effects of impurities and disorders will 
be discussed briefly in Section \ref{sec:4}. The time-reversal symmetry is
assumed throughout the paper.
Our analysis here assumes that the Hamiltonian is generic, and 
we exclude the Hamiltonians which require fine tuning of parameters. 
In other words, we exclude the cases which are vanishingly improbable
as a real material.

\section{Phase transition between the quantum spin Hall and 
insulator phases}
\label{sec:2}
As we found in Ref.~\onlinecite{Murakami07a}, 
in 2D ${\cal I}$-symmetric systems, the gap closing at 
the QSH-I 
transition occurs at TRIM $\mathbf{k}=\bm{\Gamma}_i$. This corresponds 
to the expression of the $Z_2$ topological number as a product of 
the parity eigenvalues over all the TRIMs
$\mathbf{k}=\bm{\Gamma}_i$  over the
occupied states \cite{Fu06a}; 
namely at the transition the conduction and valence 
bands with opposite parities 
exchange their parities and the $Z_2$ topological number changes.
On the other hand, for ${\cal I}$-asymmetric 2D systems,
the gap closes at $\pm 
\mathbf{k}_{0}+\bm{\Gamma}_i$ ($\mathbf{k}_0\neq 0$) by tuning some parameter.
In correspondence with this gap closing, the $Z_2$ topological number
should be expressed as an integral over the $\mathbf{k}$ space. 
This is indeed the Pfaffian expression of the $Z_2$ topological 
number \cite{Kane05a,Fu06a}.

The phase transition in 3D \cite{Fu06b,Moore06} 
can be studied similarly to 2D \cite{Murakami07a}.
The generic phase diagram in 3D is shown 
in Fig.~\ref{fig:phase-diagram-3D}(a) and is different from 2D
(Fig.~\ref{fig:phase-diagram-3D}(b)). 
In ${\cal I}$-asymmetric 3D systems, gapless phase emerges \cite{Murakami07b}, 
which 
is nonexistent in 2D. This gapless phase arises from a topological nature
of the gap-closing points in 3D. 
Namely the gap-closing points in 3D $\mathbf{k}$ space 
are monopoles and antimonopoles,
whose ``monopole charges'' are conserved.
This conservation restricts the form of the QSH-I phase transition,
as we see in this paper.

\subsection{General description of the QSH-I phase transition in 3D}
Because the QSH results from the spin-orbit coupling, the $Z_2$ topological 
number $\nu$ is $\nu=0 $ mod 2, 
when the spin-orbit coupling is zero. When we think of 
switching on the spin-orbit coupling, some may undergo a phase transition 
to the QSH phase. This phase transition changes the $Z_2$ topological number, 
which means that it is accompanied by a closing of the bulk gap.
Thus to search for candidate materials for the QSH phase, 
we consider tuning of a single parameter to drive the 
phase transition. Let us call the parameter $m$.

As we mentioned previously \cite{Murakami07b,Murakami07a},
the phase transition is 
different whether the system considered is (i) ${\cal I}$-symmetric 
or (ii) ${\cal I}$-asymmetric. 
The reason for the difference is the following.
When (i) the ${\cal I}$-symmetry is 
present, all the states are doubly 
degenerate due to Kramers theorem.
The problem is how many parameters should be tuned to 
close the gap between the conduction band and 
the valence band, which are both doubly degenerate.
This number is called a codimension, and in this case
it is five, which exceeds the number of parameters
$(\mathbf{k},m)$. 
Namely there are five independent parameters for hybridization between the 
valence and conduction bands, and unless they are finely tuned to be zero
simultaneously, the gap never close, and 
the phase transition does not occur.
 This is interpreted as
level repulsion between the valence and conduction bands.

Nevertheless, there is an exceptional case in 
(i) ${\cal I}$-symmetric systems.  
At TRIM $\mathbf{k}=\bm{\Gamma}_i$, 
all the states are classified in terms of the parity eigenvalues $(=\pm 1)$.
When the valence and conduction bands have the same parities, 
the Hamiltonian becomes \cite{Murakami07b}
\begin{equation}
{H}(\mathbf{k})=E_{0}(\mathbf{k})+\sum_{i=1}^{5}a_{i}(\mathbf{k})
\Gamma_{i},
\end{equation}
where $a_{i}$'s and $E_{0}$ are real even functions of $\mathbf{k}$.
$\Gamma_{i}$ are $4\times 4$ matrices given by
$\Gamma_{1}=1 \otimes
\tau_{x}$, $\Gamma_{2}=\sigma_{z}\otimes\tau_{y}$, 
$\Gamma_{3}=1 \otimes\tau_{z}$, 
$\Gamma_{4}=\sigma_{y}\otimes\tau_{y}$, and  
$\Gamma_{5}=\sigma_{x}\otimes\tau_{y}$, where 
$\sigma_{i}$ and $\tau_{i}$ are Pauli matrices.
The gap closes when $a_{i}(\mathbf{k})=0$ for $i=1,\cdots,5$.
It means that the codimension is five, and the gap never closes in this case.
On the other hand, 
when the valence and conduction bands have the opposite parities, 
the Hamiltonian reads \cite{Murakami07b},
\begin{equation}
{H}(\mathbf{k})=a_{0}(\mathbf{k})+a_{5}(\mathbf{k})\Gamma'_{5}+\sum_{j=1}^{4}
b^{(j)}(\mathbf{k})
\Gamma'_{j},
\end{equation}
where $a_0(\mathbf{k})$ and $a_{5}(\mathbf{k})$ are even functions of $\mathbf{k}$, and 
$b^{(j)}(\mathbf{k})$ $(j=1,2,3,4)$ are odd functions of $\mathbf{k}$. Here
$\Gamma'_{i}$  are $4\times 4$ matrices given by  
$\Gamma'_{1}=\sigma_{z}\otimes\tau_{x}$,
$\Gamma'_{2}=1 \otimes\tau_{y}$,
$\Gamma'_{3}=\sigma_{x}\otimes\tau_{x}$,
$\Gamma'_{4}=\sigma_{y}\otimes\tau_{x}$,
and $\Gamma'_{5}=1 \otimes\tau_{z}$.
In this case the gap closes only when five equations $a_{5}(\mathbf{k})=0$,
$b^{(j)}(\mathbf{k})=0$ are satisfied. 
At the TRIM, $b^{(j)}(\bm{\Gamma}_{i})$ $(j=1,2,3,4)$ identically vanish, 
and only one condition $a_{5}(\bm{\Gamma}_{i})=0$ remains to
be satisfied. This means that the codimension is reduced to one.
Thus if the valence and conduction bands have opposite parities, four 
matrix elements (out of five) 
for hybridization between the valence and conduction bands
vanish identically. The resulting codimension, i.e. the number of
parameters to be tuned for gap closing, is one. 
This is equal to the number of the tunable parameter, $m$.
(We note that in this case $\mathbf{k}$ is fixed to be $\bm{\Gamma}_i$.)
To summarize the gap can close only at $\mathbf{k}=\bm{\Gamma}_i$
for ${\cal I}$-symmetric systems. This gap-closing occurs with an 
exchange of parities between the conduction and valence bands,
as is similar to 2D.

On the other hand, 
(ii) if ${\cal I}$-symmetry is absent, the bands are not degenerate
(except for the points with $\mathbf{k}=\bm{\Gamma}_i$).
In this case the resulting codimension $D_c$ is three \cite{vonNeumann29,Herring37,Murakami07b}, 
which is smaller than that in the
inversion-symmetric case. This is because the bands are nondegenerate
and the level repulsion is less stringent.
In the symmetry classification of Wigner and Dyson \cite{Wigner,Dyson}, 
the ${\cal I}$-symmetry
breaking makes the symmetry class from symplectic ($D_c=5$) to unitary ($D_c=3$).
The codimension ($D_c=3$) 
is less than the number of parameters $(m,k_x,k_y,k_z)$.
Thus the gap can close by tuning a parameter $m$.

\subsection{Monopole-antimonopole pair creation and annihilation 
in $\mathbf{k}$ space}
Henceforth we focus on (ii) the ${\cal I}$-asymmetric systems. 
Our theory is based on the physics of gauge
field in $\mathbf{k}$-space
\cite{Berry84, Volovik}. 
When $\alpha$-th band is degenerate with another band at an 
isolated point $\mathbf{k}$,
such point is associated with a monopole for the gauge 
field in $\mathbf{k}$-space. The gauge field $\mathbf{A}_{\alpha}
(\mathbf{k})$ and the 
corresponding field strength $\mathbf{B}_{\alpha}(\mathbf{k})$ are defined as
\begin{align}
&\mathbf{A}_{\alpha}
(\mathbf{k})=-i\langle \psi_{\alpha}(\mathbf{k})|\nabla_{\mathbf{k}}|
\psi_{\alpha}(\mathbf{k})\rangle,\label{eq:A}\\
&\mathbf{B}_{\alpha}
(\mathbf{k})=\nabla_{\mathbf{k}}\times\mathbf{A}_{\alpha}(\mathbf{k}),
\label{eq:B}
\end{align}
The monopole density is defined as 
\begin{equation}
\rho_{\alpha}
(\mathbf{k})=\frac{1}{2\pi}\nabla_{\mathbf{k}}\cdot\mathbf{B}_{\alpha}
(\mathbf{k})
\end{equation}
Though at first sight $\rho_{\alpha}(\mathbf{k})$ vanishes 
identically, 
because $\nabla_{\mathbf{k}}\cdot (\nabla_{\mathbf{k}}\times \ )=0$, 
it is not true. In some cases where the $\alpha$-th band 
touches with another band at some
$\mathbf{k}$-point, the wavenumber cannot be chosen as 
a single continuous function for the whole Brillouin zone. 
In such case the Brillouin zone should be patched with more than 
one continuous wavefunctions \cite{Kohmoto85}, 
as is similar to the vector potential of the Dirac monopole \cite{WuYang}. 
This allows a $\delta$-function singularity of 
$\rho(\mathbf{k})$ at the band touching.  
As a result the monopole density has the form $\rho(\mathbf{k})=\sum_{l}q_{l}
\delta(\mathbf{k}-\mathbf{k}_{l})$ where $q_{l}$ is an integer called
a monopole charge.
Even when we vary the system by changing a parameter continuously, 
the monopole charge is conserved, because it is quantized. 
The only chance for the monopole charge to change 
is to create or to annihilate a pair of a monopole ($q_{l}=1$) and
an antimonopole ($q_{l'}=-1$).  
More detailed formulation is in Appendix A.

In the present case, we restrict ourselves to 
time-reversal symmetric cases, where we have
\begin{equation}
\mathbf{B}_{\alpha}(\mathbf{k})=-\mathbf{B}_{\bar{\alpha}}(-\mathbf{k}), \ 
\rho_{\alpha}(\mathbf{k})=\rho_{\bar{\alpha}}(-\mathbf{k}), 
\end{equation}
where $\bar{\alpha}$ is the label which is a time-reversed label from $\alpha$.
It means that the monopoles distribute symmetrically 
with respect to the origin.

At the phase transition, 
the gap closes between a single valence band and a 
single conduction band  at $\mathbf{k}=\mathbf{k}_{0}$.
Instead of considering a general Hamiltonian 
it is sufficient and much simpler 
to consider a $2\times 2$ matrix $H(\mathbf{k},m)$.
Here we introduce an external parameter $m$, which controls the phase 
transition.
We note that the following discussion on $2\times 2$
Hamiltonian 
is easily generalized to an arbitrary Hamiltonian.
The 2$\times$2  Hamiltonian $H(\mathbf{k},m)$ is expanded as
\begin{equation}
H(\mathbf{k},m)=a_0(\mathbf{k},m)+\sum_{i=1}^{3}a_{i}(\mathbf{k},m)
\sigma_i,
\end{equation}
where $\sigma_i$ ($i=1,2,3$) are the Pauli matrices. 
The gap closes when the two eigenvalues are identical, i.e.
when the three conditions
$a_{i}(\mathbf{k},m)$=0 ($i=1,2,3$) are satisfied. Therefore, in 
general, the gap-closing point in the $(k_x,k_y,k_z,m)$-hyperspace
forms a curve, which we call a ``string''.
Generally, this string $C$ occupies a finite region in 
$m$-direction; namely, it is vanishingly improbable to 
lie in a single value of $m$. 
When we cut the string $C$ at some value of $m$,
the intersections in the $\mathbf{k}$-space are the 
points where the gap closes, namely, the monopoles and antimonopoles. 
Thus the string $C$ is the trajectory of the monopoles and 
antimonopoles. 
Because the monopole charge is conserved,
the monopoles and antimonopoles are created and annihilated 
only in pairs, which means that 
the trajectory $C$ of the monopoles and 
antimonopoles forms a closed loop in the $(\mathbf{k},m)$ space 
(see Fig.~\ref{fig:string} in Appendix A).
Namely, the string $C$ has no end point, 
because an end point of $C$ 
would violate the conservation
of monopole charge.

Henceforth we describe how the phase transition occurs, thereby 
opening a gap. 
We consider a situation where one side of $m$, e.g. $m<m_0$
is gapped while the other side of $m$, e.g. $m>m_0$ is
gapless. This means that the string $C$ exists only in the $m>m_0$ region. 
In other words, we consider an extremum of the string
$C$. 
We pick up a gap-closing point $(\mathbf{k},m)=(\mathbf{k}_{0},m_{0})$,
(i.e. $\mathbf{a}(\mathbf{k}_{0},m_{0})=0$) 
and consider the vicinity of this point.
We investigate conditions for the point $(\mathbf{k}_{0},m_{0})$ to become 
an extremum of the string $C$.
We expand the coefficients to the linear order 
\begin{equation}
a_{i}(\mathbf{k},m)=
\sum_{j}M_{ij}\Delta k_{j}
+N_{i}\Delta m,
\end{equation}
or in a matrix form
\begin{equation}
\mathbf{a}(\mathbf{k},m)=
M\Delta \mathbf{k}
+\Delta m\mathbf{N},
\label{eq:a-matrix}
\end{equation}
where $\Delta k_{j}=k_{j}-k_{0j}$, $\Delta m=m-m_{0}$,
$M_{ij}=\left.\frac{\partial a_{i}}{\partial k_{j}}
\right|_{0}$ and 
$N_i
=\left.\frac{\partial a_{i}}{\partial m}\right|_{0}$.
If the determinant of the 
matrix $M$ does not vanish, the gap-closing condition, 
$\mathbf{a}=(a_1, a_2, a_3)=0$ gives
\begin{equation}
\Delta\mathbf{k}=-M^{-1}\mathbf{N}\Delta m.
\end{equation}
It means that a gap-closing point moves as the parameter $m$ 
changes, and it exists on the both sides of $m=m_{0}$.
It is not the case of our interest.
Therefore we conclude
\begin{equation}
\mathrm{det}M\equiv \mathrm{det}_{(i,j)}
\left.\frac{\partial a_{i}}{\partial k_{j}}
\right|_{0}=0,
\label{eq:detM}
\end{equation}
which is imposed in addition to $a_{i}=0$. 
Thus there are four conditions in total, which give a set
of gap-closing points $(\mathbf{k}_{0},m_{0})$ located at an extremum 
of the string $C$. 

\begin{figure}
\includegraphics[scale=0.37]{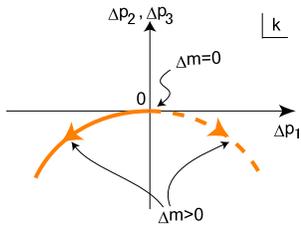}
\caption{(Color online) 
Trajectory for the gapless points in the $\mathbf{k}$ space.
At $m=m_0$ a monopole-antimonopole pair is created, and they run to the 
opposite directions when $m$ is changed. 
We assume $(Q^{-1}\mathbf{N})_{1}<0$, in which 
gapless points exist only for $\Delta m\geq 0$. }
\label{fig:trajectory-mono}\end{figure}

We now calculate behaviors of the system in the vicinity 
of $(\mathbf{k}_{0},m_{0})$. 
If Eq.~(\ref{eq:detM}) holds, the matrix $M$ has a normalized eigenvector 
$\mathbf{n}_{1}$ with
null eigenvalue: $M\mathbf{n}_{1}=0$.
From $\mathbf{n}_{1}$ we consider two additional unit 
vectors $\mathbf{n}_{\alpha}$ ($\alpha=2,3$) to form an
orthonormal basis $\left\{\mathbf{n}_{1},\mathbf{n}_{2},\mathbf{n}_{3}
\right\}$.
We adopt this basis for the $\mathbf{k}$ space;
\begin{equation}
\Delta\mathbf{k}=U\Delta\mathbf{p}\equiv
\left(\mathbf{n}_{1},\mathbf{n}_{2},\mathbf{n}_{3}
\right)\left(\begin{array}{c}\Delta p_{1}\\ \Delta p_{2} \\ \Delta p_{3}
\end{array}
\right).
\label{eq:Deltakunitary}
\end{equation}
Namely, $(\Delta p_1,\Delta p_2,\Delta p_3)$ is a 
coordinate rotated from $(\Delta k_1,\Delta k_2,\Delta k_3)$.
From (\ref{eq:a-matrix}) and (\ref{eq:Deltakunitary}) to the linear order in 
$\Delta\mathbf{k}$ and $m$,
we have
\begin{equation}
\mathbf{a}=\Delta p_{2}\mathbf{u}_{2}+
\Delta p_{3}\mathbf{u}_{3}+\Delta m \mathbf{N},
\end{equation}
where $\mathbf{u}_{i}=M\mathbf{n}_{i}$ ($i=2,3$). 
Up to this order, 
the gap closing condition, $\mathbf{a}=0$, has no nontrivial 
solution  in general,
because the three vectors $\mathbf{u}_{2}$, 
$\mathbf{u}_{3}$, 
$\mathbf{N}$ are generally linearly independent.
It is not the case of our interest. 
Thus we have to include the next order in $\Delta \mathbf{k}$ and
$\Delta m$, to see
whether the gap closes for $\Delta m\neq 0$;
\begin{align}
&\mathbf{a}=\Delta m \mathbf{N}+\Delta p_{2}\mathbf{u}_{2}+
\Delta p_{3}\mathbf{u}_{3}+\sum_{i,j=1,2,3,i\leq j}\mathbf{u}_{ij}\Delta
p_{i}\Delta p_{j}\nonumber \\
&\ \ +\sum_{i=1}^{3}\tilde{\mathbf{u}}_{i}\Delta m\Delta p_{i}
+\mathbf{u}(\Delta m)^2,
\label{eq:a}
\end{align}
where $\mathbf{u}_{ij}$, $\tilde{\mathbf{u}}_{i}$ and $\mathbf{u}$
are vectors.
We first look at the gap-closing point $\mathbf{a}=0$.
We put $\Delta m \propto \lambda$ 
where $\lambda$ is small, and investigate the order of $\lambda$ for 
each term. As we have seen, if $\Delta p_{i}$ ($i=1,2,3$) are
of the order $\lambda$, the gap-closing condition has no nontrivial 
solution.
The reason is that 
the right-hand side of Eq.~(\ref{eq:a}) has no term linear in $\Delta p_{1}$. 
Hence we have to consider the quadratic term in $\Delta p_{1}$, 
for which we put $\Delta p_{1}\propto \lambda^{1/2}$. Then up to $O(\lambda)$
we have
\begin{equation}
\Delta m \mathbf{N}+\Delta p_{2}\mathbf{u}_{2}+
\Delta p_{3}\mathbf{u}_{3}+
(\Delta p_{1})^{2}\mathbf{u}_{11}=0.
\end{equation}
The solution is given by 
\begin{equation}
\left(
\begin{array}{c}
(\Delta p_{1})^{2} \\ \Delta p_{2}\\ \Delta p_{3}
\end{array}\right)=-\Delta m (Q^{-1}\mathbf{N}), \ 
Q=\left(\mathbf{u}_{11},\mathbf{u}_{2},\mathbf{u}_{3}\right).
\end{equation}
Because $\Delta p_{1}^{2}\geq 0$, 
the solution for this exists only when $\Delta m$ has the same 
sign with $-(Q^{-1}\mathbf{N})_{1}$. This means that on one side
of $\Delta m=0$ the system is gapped, while on the other side the system 
has gapless points,
\begin{align}
&\Delta p_{1}=\pm\sqrt{-(Q^{-1}\mathbf{N})_{1}\Delta m},\\
&\Delta p_{2}=-(Q^{-1}\mathbf{N})_{2}\Delta m,\\
&\Delta p_{3}=-(Q^{-1}\mathbf{N})_{3}\Delta m.
\end{align}
Thus when $\Delta m$ is changed across zero, 
monopole-antimonopole pairs are created
and dissociate along the $\Delta p_{1}$ direction.
The trajectory of the gapless points is as shown in 
Fig.~\ref{fig:trajectory-mono}.
This is exactly the case of our pursuit: the point of an 
extremum of the gap-closing. 
Thus we have shown that for given Hamiltonian
with broken ${\cal I}$-symmetry, such point exists in general, 
and the behavior of the gapless points in the vicinity of this pair
creation or annihilation is described.

Next we consider the dispersion around the gap-closing point. 
Around the monopole the dispersion is linear in $\mathbf{k}$. 
It is, however, not the case around the point of monopole-antimonopole
pair creation. 
From Eq.~(\ref{eq:a}), we can derive $\mathbf{k}$ dependence of the gap
at $\Delta m=0$ (pair creation or annihilation).
The gap is given by $E_{\mathrm{g}}=2|\mathbf{a}|$. Hence, for $\Delta m=0$. 
The gap behaves as
\begin{equation}
E_{\mathrm{g}}\propto \Delta p_{2}, \Delta p_{3}, (\Delta p_{1})^{2}
\end{equation}
Thus the dispersion along the $\Delta p_{1}$ direction is quadratic,
while that along the $\Delta p_{2}$ and $\Delta p_{3}$ directions is
linear. The $\Delta p_{1}$-direction is the direction for the 
monopole-antimonopole pair to dissociate.

\begin{figure}
\includegraphics[scale=0.37]{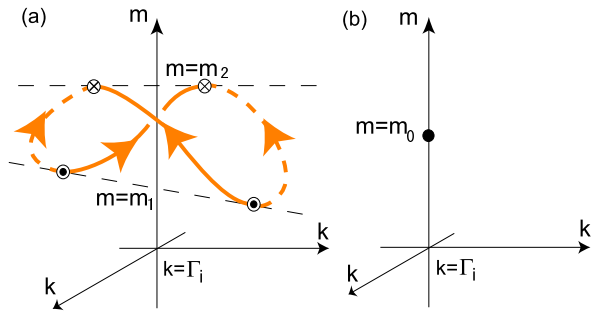}
\caption{(Color online) 
Trajectory of the gapless points for (a) inversion-asymmetric and 
(b) symmetric systems. For (b) inversion-symmetric systems, the gapless
point is located at $\mathbf{k}=\bm{\Gamma}_i$, and is an isolated point in the
$m$-$\mathbf{k}$ space. Only 
at $m=m_0$ the system is gapless. 
For (a) inversion-asymmetric systems, on the other 
hand, the gapless points are created in monopole-antimonopole pairs
at $m=m_1$, and
move in $\mathbf{k}$-space as $m$ is varied. The system opens a gap only 
by pair annihilation of these gapless points at $m=m_2$. }
\label{fig:monopole}\end{figure}

\subsection{Change in the $Z_2$ topological numbers}
According to Ref.~\onlinecite{Fu06b}, 
the $Z_2$ topological numbers $\nu_j$ ($j=0,1,2,3$) in 3D 
are defined
as 
\begin{align}
&(-1)^{\nu_0}=\prod_{n_j=0,1}\delta_{n_1n_2n_3},\label{nu0}\\
&(-1)^{\nu_{i=1,2,3}}=\prod_{n_{j\neq i}=0,1;n_i=1}\delta_{n_1n_2n_3},
\label{nui}\end{align}
where 
\begin{align}
\delta_{i}=\sqrt{\mathrm{det}[w(\bm{\Gamma}_i)]}/
\mathrm{Pf}[w(\bm{\Gamma}_i)]=\pm 1.
\end{align}
Here $w_{nm}=\langle u_{m,\mathbf{-k}}|\Theta| u_{n,\mathbf{k}} \rangle$
, where $\Theta$ is the time-reversal operator, 
and $u_{m,\mathbf{k}}$ is the periodic part of the Bloch wavefunction.
Each phase is expressed as $\nu_0; (\nu_1\nu_2\nu_3)$, which 
distinguishes 16 phases.
Because among $\nu_i$, $\nu_0$ is the only topological number which 
is robust against disorder, the phases are mainly 
classified by $\nu_0$.
When $\nu_{0}$ is odd the phase is called as the strong topological 
insulator (STI), while if it is even it is called the weak topological 
insulator (WTI). The STI and WTI correspond to the QSH and I phases, 
respectively.
The other indices $\nu_1$, $\nu_2$, and $\nu_3$ are used
to distinguish various phases in the STI or WTI phases,
and each phase can be associated with a mod 2 reciprocal lattice vector
$\mathbf{G}_{\nu_1\nu_2\nu_3}=\nu_1\mathbf{b}_{1}
+\nu_2\mathbf{b}_{2}
+\nu_3\mathbf{b}_{3}$, as was proposed in Ref.~\onlinecite{Fu06b}.

We can relate the shape of the trajectory (``loop'') 
of the gapless points with
the change in the topological number. 
To see this, 
we note the following. From Eqs.~(\ref{nu0}) and (\ref{nui}), 
the $Z_2$ topological numbers in 3D are defined on 
planes $S_{i}^{(n_{i})}$: $\mathbf{k}\cdot \mathbf{a}_{i}=
\pi n_i$ with $n_i=0,1$
 in the Brillouin zone as
\begin{equation}
\nu_0\equiv\prod_{\bm{\Gamma}_j}\delta_{j},\ \ 
\nu_i\equiv\prod_{\bm{\Gamma}_j\in S_{i}^{(1)}}\delta_{j},
\end{equation}
which are gauge invariant and have the values $\pm 1$. 

Let us take the plane $S_1^{(1)}$ for example. This affects the numbers
$\nu_0$ and $\nu_1$. 
Based on the theory on homotopy characterization of the 2D QSH phase
\cite{Moore06}, one can show the following. 
An intersection of the loops with the plane $S_1^{(1)}$ forms a set of isolated
points. They are symmetric with respect to $\mathbf{k}=\bm{\Gamma}_{(100)}$, 
and 
the number of points is even. When the number of the points is
$2(2N+1)$ ($N$: integer),
 then it accompanies the change in the $Z_2$ topological number 
$\nu_0$ and $\nu_1$. 
Otherwise, when the number is $4N$ ($N$: integer), then it does 
not accompany the change in the $Z_2$ topological numbers.
To show this we note the result in Ref.~\onlinecite{Moore06}:  
in 2D the $Z_2$ topological number $\nu$ is equal (modulo 2) to an 
integral of the Berry curvature inside a half of the Brillouin 
zone plus an extra term coming from ``contraction'' of the 
Brillouin zone \cite{Moore06}.
We can regard the slice of the 3D Brillouin zone by $S_1^{(1)}$ as
a 2D Brillouin zone \cite{Moore06}, which we call $D_1^{(1)}$.
By this identification we treat the 3D $Z_2$ topological numbers  
in the same way as in 2D.
When the loop intersects the half of the Brillouin 
zone $D_1^{(1)}$ (within $S_1^{(1)}$) once, 
it means that at some $m$ the monopole 
passes through the half of the Brillouin zone $D_1^{(1)}$, and changes the 
integral of the Berry curvature by unity. Thus it changes $\nu_0$ and
$\nu_1$ as 
$\nu_0 \rightarrow \nu'_{0}\equiv \nu_0+1$ (mod 2), 
$\nu_1 \rightarrow \nu'_{1}\equiv \nu_1+1$ (mod 2).
Therefore if $2N+1$ intersections occur within the half of $D_1^{(1)}$, the $Z_2$ topological numbers
$\nu_0$ and $\nu_1$ 
changes (odd $\leftrightarrow$ even), 
while if $2N$ intersections occur, $\nu_0$ and
$\nu_1$ are unchanged.
This completes the proof that 
when a number of intersections between
the loops and the plane $S_1^{(1)}$ is
$2(2N+1)$ ($N$: integer), it accompanies 
the change in the $Z_2$ topological numbers,
 whereas $4N$ intersections involve no change in the 
$Z_2$ topological numbers.

\begin{figure}
\includegraphics[scale=0.37]{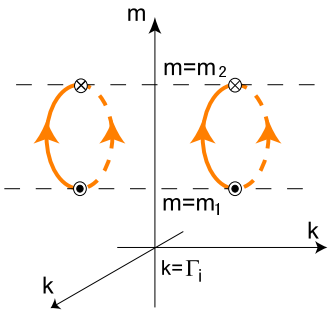}
\caption{(Color online) 
Trajectory of the gapless points for inversion-asymmetric systems,
but without the phase transition.}
\label{fig:monopole-notransition}\end{figure}

For further investigation, we consider the following example. 
Suppose we consider simultaneous pair creations at $m=m_1$, and 
$\mathbf{k}=\pm\mathbf{k}_{0}+\bm{\Gamma}_i$. 
As a result we have two monopoles and two antimonopoles. 
Eventually these will be annihilated at 
$\mathbf{k}=\pm\mathbf{k}'_{0}+\bm{\Gamma}_i$.
Then there are two possible cases: 
pair annihilation occurs (A) with changing partners
(Fig.~\ref{fig:monopole}(a)) and (B) without changing the partners
(Fig.~\ref{fig:monopole-notransition}).
We can show that (A) 
changes the $Z_2$ topological numbers while (B) does not.
One can see the reason in two different ways. 
One way to see the difference 
is to consider an intersection 
of the ``loops''  with the planes such as $S_{i}^{(n_i)}$. 
In this case, on half of the Brillouin zone the number of intersection 
points is necessarily even. 

The other way to see this difference between (A) and (B) 
is to consider switching on a perturbation which 
restores the ${\cal I}$-symmetry.
One may wonder whether it is 
possible to restore the 
${\cal I}$-symmetry without encountering a phase transition.
When ${\cal I}$-symmetry-breaking term in the 
Hamiltonian is sufficiently small, it is possible, 
whereas in generic systems we cannot prove that it is possible.
Therefore, in the following we assume that it is 
possible to restore the 
${\cal I}$-symmetry without encountering a phase transition.
In this case, the gapless loop eventually reduces to a point 
 in (A) (Fig.~\ref{fig:monopole}(a)). Meanwhile in the case (B)
(Fig.~\ref{fig:monopole-notransition}) the loops cannot reduce  to
a point in the ${\cal I}$-symmetric limit.
In the case (B), by restoring the ${\cal I}$-symmetry, 
each of the two loops seems to shrink to a point 
($\neq \bm{\Gamma}_{i}$); this, however, is impossible
because in ${\cal I}$-symmetric case the gap closing does
not occur at $\mathbf{k}\neq \bm{\Gamma}_{i}$ due to the large
codimension $(=5)$.
Thus in (B), the perturbation can get rid of the gapless points
completely from the $(\mathbf{k},m)$ space, i.e. the two phases
of both sides are identical.

From these arguments one can see that the change of $\nu_0$ is equal
to the number of loops in $m$-$\mathbf{k}$ space, modulo 2. 
In the case (A) (Fig.~\ref{fig:monopole}(a)), there is a single loop, 
and the change of $\nu_0$ is one, while in the case (B) 
(Fig.~\ref{fig:monopole-notransition}), the number of loop is two,
and $\nu_0$ is unchanged.

In the analysis in this section, we assumed that as the external 
parameters are changed, the Hamiltonian
and the band structure change continuously.
Namely, we assumed an absence of first-order transitions.
Whether or not first-order transitions happen 
depends on details of the system and is not solely determined 
by system symmetries and topological order. 
Therefore, if first-order transitions are taken into account, 
it is no longer possible to discuss universal properties.
For example, when first-order transitions are allowed, the two
phases, QSH and I phases, can transit to each other via first-order 
transition without closing 
a bulk gap. This situation is realized in the model in Ref.~\onlinecite{Raghu08}, 
where the topological QSH phase is realized as a phase with 
spontaneously broken symmetry.

\section{Example: 3D Fu-Kane-Mele model}
\label{sec:3}
To confirm the topological discussion in the previous section, 
we take the 3D model proposed by 
Fu, Kane and Mele \cite{Fu06b} on a diamond lattice as an 
example. This model
shows a transition between STI and WTI.
The model is written as 
\begin{equation}
H=t\sum_{\langle ij\rangle}c_{i}^{\dagger}c_{j}
+i(8\lambda_{\mathrm{SO}}/a^{2})\sum_{\langle\langle
ij\rangle\rangle} c_{i}^{\dagger}\mathbf{s}\cdot
(\mathbf{d}_{ij}^{1}\times \mathbf{d}_{ij}^{2})c_{j}.
\end{equation}
Here $a$ is the size of the cubic unit cell, $t$ represents the 
hopping, and $\lambda_{\mathrm{SO}}$ represents the spin-orbit coupling.
The first term represents the nearest  neighbor hopping,
and the second term is a spin-dependent hopping 
to the next nearest neighbor sites. $\mathbf{d}_{ij}^{1}$ and 
$\mathbf{d}_{ij}^{2}$ are
the vectors 
for 
the two nearest neighbor bonds included in the next-nearest-neighbor hopping.

\begin{figure}
\includegraphics[scale=0.37]{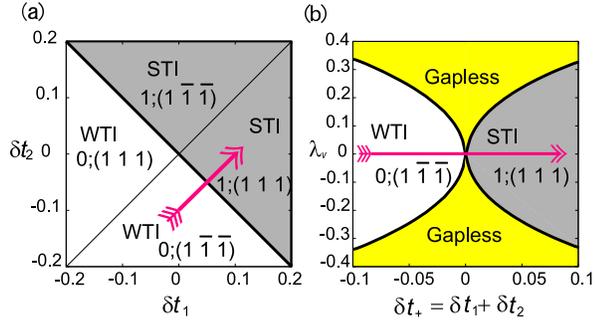}
\caption{(Color online) 
Phase diagrams for the Fu-Kane-Mele model with $\delta t_3=0$,
$\delta t_4=0$. $t_1$ and $t_2$ are the bonds along the [111] and 
[$1\bar{1}\bar{1}$] directions. 
We put $\lambda_{\mathrm{SO}}=0.1t$. The axes are in the unit
of $t$. (a) The phase diagram in 
$\delta t_1$-$\delta t_2$ plane obtained in Ref.~\onlinecite{Fu06b}.
$\lambda_v$ is set as zero.
Each phase is indexed by cubic Miller indices, following 
Ref.~\onlinecite{Fu06b}.
(b) The phase 
diagram in the $\delta t_{+}$-$\lambda_v$ plane.
$\lambda_v$ is newly introduced into the Fu-Kane-Mele model. Here 
$\delta t_{+}=\delta t_{1}+\delta t_{2}$,
while we fix $\delta t_{-}=\delta t_{1}-\delta t_{2}=0.1t$. 
The arrows (red) in (a) and (b) correspond to the identical
change in parameters. }
\label{fig:phase-dia1}
\end{figure}

This four-band model is ${\cal I}$-reversal and time-reversal symmetric. 
It means that every eigenstate is doubly degenerate by the Kramers theorem.
The doubly-degenerate conduction and the valence bands
touch at the three $X$ points, $X^r=(2\pi/a) \hat{r}$ $(r=x,y,z)$,
and therefore the bulk gap vanishes.
To consider the phases with a bulk gap,
suppose one changes the nearest-neighbor hopping to be different 
for the four directions
of nearest neighbor bonds $t_i$ ($i=1,2,3,4$) \cite{Fu06b}.
The system then opens a gap between the two doubly-degenerate bands,
while the ${\cal I}$-reversal and time-reversal symmetries are preserved.
In Ref.~\onlinecite{Fu06b}, 
the phase boundary is studied when the hopping is changed 
slightly from the identical value: $t_{i}=t+\delta t_{i}$. 
When we set $\delta t_{3}=0=\delta t_{4}$, the phase diagram is as shown 
in Fig.~\ref{fig:phase-dia1}(a) as a function of $\delta t_{1}$ and 
$\delta t_{2}$ as obtained 
in Ref.~\onlinecite{Fu06b}. 
Four phases meet at $\delta t_{1}=0=\delta t_{2}$, where
the system becomes gapless.  

The problem of our current interest is how the phase boundary between 
the WTI and the STI changes
when the ${\cal I}$-symmetry is broken.
In the present model, the simplest way to break ${\cal I}$-symmetry is
to introduce an alternating on-site energy $\lambda_v$, like in 
the 2D Kane-Mele (KM) model on the
honeycomb lattice \cite{Kane05a}. In the present 3D case, 
the alternating on-site energy reduces the system to be 
similar to the zincblende structure as in GaAs.

By introducing $\lambda_v$, the symmetry of the system is lowered, 
and an analytic calculation of the phase transition becomes much harder.
In the present case, however, with a procedure 
explained in Appendix B,
we can calculate how the WTI-STI phase transition changes by 
breaking the ${\cal I}$-symmetry by the $\lambda_v$ term.
In the ${\cal I}$-symmetric ($\lambda_{v}=0$) case, 
from the phase diagram (Fig.~\ref{fig:phase-dia1}(a)) 
we consider $\delta t_{+}=\delta t_{1}+\delta t_{2}$ as a parameter
$m$ driving the phase transition, while $\delta t_{1}-\delta t_{2}$ is
fixed to be a nonzero value, for example, 
$\delta t_{1}-\delta t_{2}=0.1t$. This corresponds to
the red arrow in Fig.~\ref{fig:phase-dia1}(a).
The phase diagram in the $\delta t_{+}$-$\lambda_{v}$ plane is as given 
by Fig.~\ref{fig:phase-dia1}(b). 
When the ${\cal I}$-symmetry is broken ($\lambda_{v}\neq 0$),
the gapless region appears in the phase diagram.
This confirms our theory in the previous section.
As we explicitly show at the end of Appendix B, even
when the model parameters are changed perturbatively, the gapless
points move but never disappear. In this sense the gapless phase 
is stable.

\begin{figure}[htb]
\includegraphics[scale=0.35]{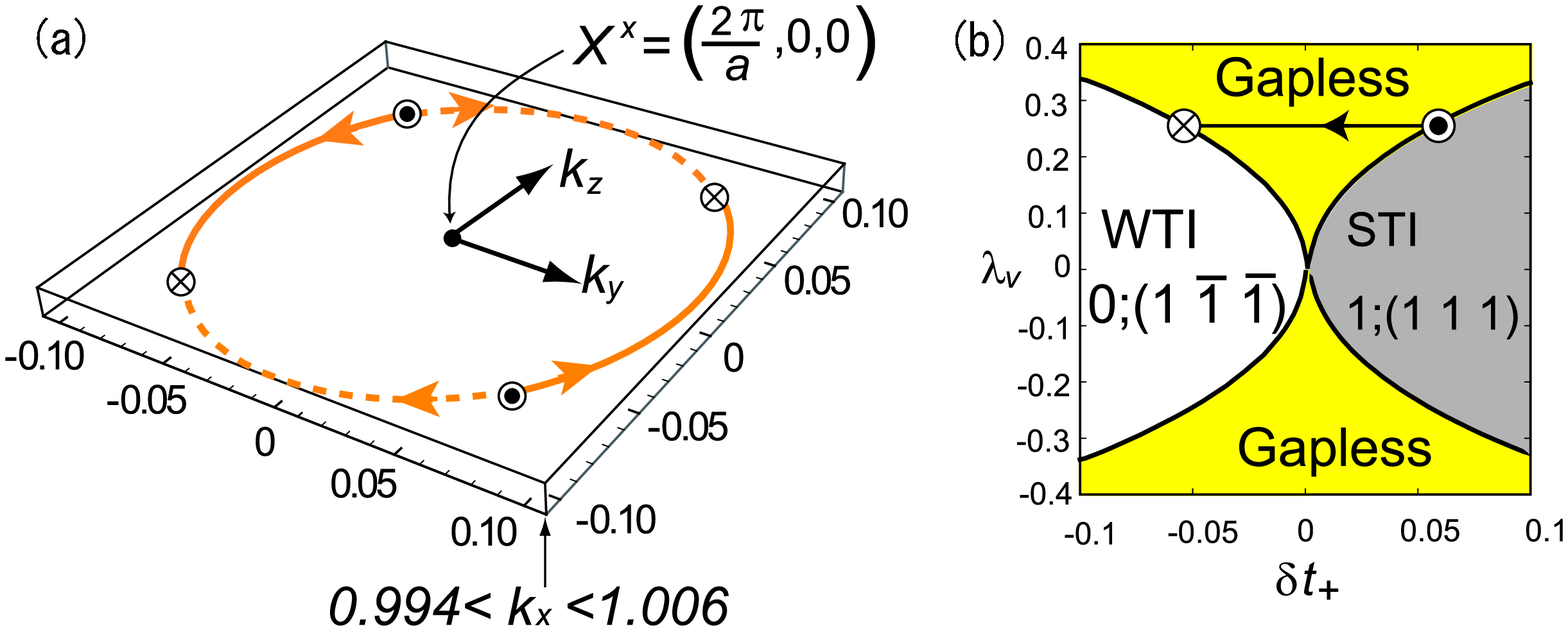}
\caption{(Color online)
(a) Trajectory of the gapless points in $\mathbf{k}$ space,
as we change the parameter $\delta t_{+}$ with $\lambda_v$ fixed.
The wavenumber $\mathbf{k}$ is shown in the unit of $(2\pi/a)$.
The solid and broken curves are the trajectories for 
the monopoles and antimonopoles, respectively.
This corresponds to the arrow in the phase diagram in (b).}
\label{fig:trajectory-mono2}
\end{figure}

To confirm our theory further, 
we calculate the trajectory (``string'') 
of the gapless points in $\mathbf{k}$ space.
As the parameter $\delta t_{+}$ is
changed along the arrow in Fig.~\ref{fig:trajectory-mono2}(b), the
gapless points move in $\mathbf{k}$ space, as shown in 
Fig.~\ref{fig:trajectory-mono2}(a). As a whole, the trajectory 
is almost circular (but not exactly) in the $\mathbf{k}$ space, around the 
$X^x$ point. Note that when we gradually decrease $\lambda_v$, the 
trajectory is reduced to the $X^x$ point. 

The change in the $Z_2$ topological numbers can be
seen by counting the intersection points between the trajectory
and the planes $S_i^{(n_i)}$.
We choose
the reciprocal lattice vectors as $\mathbf{b}_{1}=\frac{2\pi}{a}(-1,1,1)$,
$\mathbf{b}_{2}=\frac{2\pi}{a}(1,-1,1)$,
and $\mathbf{b}_{3}=\frac{2\pi}{a}(1,1,-1)$.
The $X^x$ point ($=(\mathbf{b}_{2}+\mathbf{b}_{3})/2=\bm{\Gamma}_{(011)}$) 
then lies 
on the planes $S_1^{(0)}$, $S_2^{(1)}$,
and $S_3^{(1)}$, and 
the trajectory intersects these planes twice.
The trajectory
does not 
intersect the planes $S_1^{(1)}$, $S_2^{(0)}$, $S_3^{(0)}$. 
This means that among the $Z_2$ topological 
numbers, only $\nu_0$, $\nu_2$ and $\nu_3$ changes.

\begin{figure}[htb]
\includegraphics[scale=0.33]{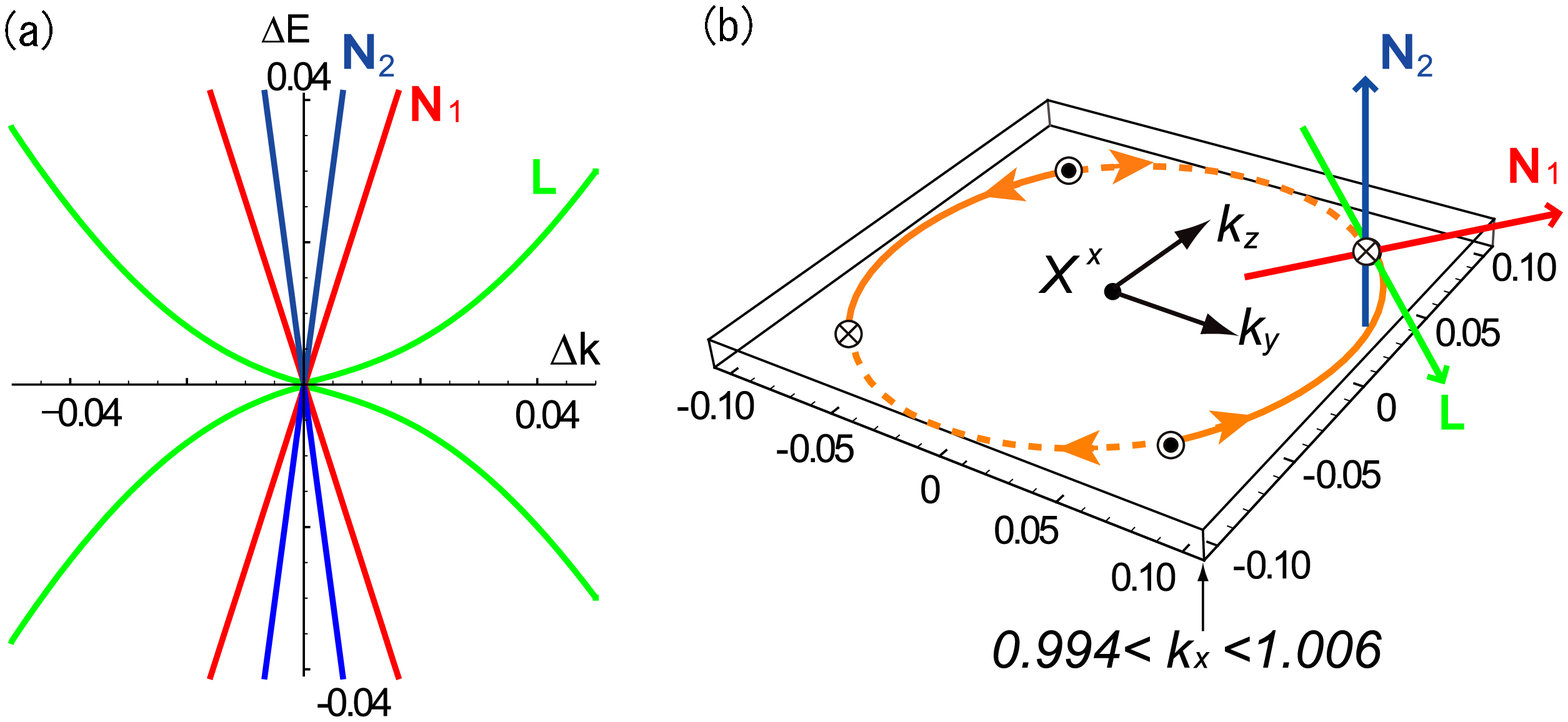}
\caption{(Color online) 
(a) Dispersion around the gapless points
when the monopole-antimonopole pairs annihilate,
for the 
Fu-Kane-Mele model.
The direction $\mathbf{L}$ is along the trajectory of the 
monopole and the antimonopole, while $\mathbf{N}_1$ and 
$\mathbf{N}_2$ are perpendicular
to $\mathbf{L}$. 
These directions are shown in (b) 
in the trajectory for the gapless point in the $\mathbf{k}$ space.
Energy is shown in the unit $t$, and the wavenumber is in the unit $2\pi/a$. }
\label{fig:dispersion}
\end{figure}

In the previous section we predicted that 
the dispersion at the pair creation and annihilation is
anisotropic. To check this, we pick up a point of pair annihilation
(Fig.~\ref{fig:dispersion}), and calculate the dispersion 
around this point. Indeed, along the directions $\mathbf{N}_{1}$
and $\mathbf{N}_{2}$ the dispersion is linear; meanwhile, 
along the direction 
$\mathbf{L}$ which is tangential to 
the trajectory of the monopole and the antimonopole, 
the dispersion is much softer and quadratic. This 
agrees with the theory in the previous section.

\section{Conclusions and Discussions}
\label{sec:4}
In this paper, we described the generic phase diagrams involving
the quantum spin Hall and insulator phases,
by tuning an external parameter in 2D and 3D.
In ${\cal I}$-asymmetric 3D systems, there lies a finite region 
of the gapless phase in the phase diagram. 
This was checked for the Fu-Kane-Mele model.
We described the phase transition in terms of the motion of the 
gap-closing points (i.e. monopoles and antimonopoles) in $\mathbf{k}$ 
space. The gapless phase in the ${\cal I}$-asymmetric 3D system 
originates from the conservation of ``monopole charge''

We also studied the dispersion around the gapless points. 
In general the energy dispersion is linear around the gapless points.
Meanwhile, around the gapless points at monopole-antimonopole 
pair creation/annihilation, the dispersion becomes quadratic in 
one direction of $\mathbf{k}$, which is tangential to the 
monopole trajectory. 

It is interesting to compare the physics of gap-closing in
the QSH phase and that in the quantum Hall (QH) phase.
In the QH phase the bulk is gapped, and is characterized by the
Chern number, by which this phase is distinct from 
an ordinary insulator. 
For the QH phase, the behavior of monopoles and antimonopoles are
quite similar to that presented so far in this paper.
This point can be illustrated 
by using a 2D model of the quantum (anomalous) Hall 
effect proposed by Haldane \cite{Haldane88}.
The phase diagram is shown in 
Ref.~\onlinecite{Haldane88}, where the transition between 
the QSH and the insulator phases occurs
by changing the model parameter $M$. This phase transition 
occurs when the gap closes only at one wavenumber in the Brillouin zone,
which is similar to the case of the 2D QSH phase \cite{Murakami07a}
In 3D, the gapless phase is expected to 
occur between the QH phase and 
the ordinary insulator phase (and also between two QH phases 
with different Chern numbers).
It is because the symmetry class is unitary, and 
it is similar to the QSH phase without ${\cal I}$-symmetry.
The 3D QH phase is
characterized by three Chern numbers \cite{Montambaux90}. 
Therefore, for the transition between 
phases with different sets of Chern numbers, a gapless phase is
expected to appear in between. This gapless phase is described by a gapless 
loop $C$ in the $(m,\mathbf{k})$ space, as in the 3D QSH case.
The topology of the loop $C$ relative to the crystallographic directions 
determines the change in the three Chern numbers at the transition.
By comparing the cases of the QSH phase and the QH phase,
they are different in the following aspects. First, the QH 
effect is without time-reversal symmetry. This makes the physics 
at  $\mathbf{k}$ and $-\mathbf{k}$ independent, in contrast with 
the QSH case. Therefore, the presence or absence of the ${\cal I}$-symmetry
is inessential in the QH system.
Second, because the QH system usually requires a strong magnetic
field, and the motion parallel to the magnetic field is usually gapless, 
the QH phase in 3D is not easily realized in real systems.
Meanwhile there is no such constraint for the QSH phase, and 3D QSH phase
is easily realized. Thus the 3D gapless phase due to the topological nature 
of monopoles is more realistic in the 
QSH phase than in the quantum Hall phase.

When disorder is introduced in the system, the phase transition 
in three dimensions will have a gapless region as a function 
of the external parameter $m$. 
In 3D, only for ${\cal I}$-symmetric systems, 
the disorder effect on the phase transition between the QSH and 
I phases has been studied \cite{Shindou08}, while the 
${\cal I}$-asymmetric systems are left to be analyzed.
By regarding the whole system as one unit cell (``supercell''),
the similar discussion holds.
We impose the periodic boundary conditions, but with allowing 
additional phase twisting for the individual directions
as $\theta_x$, $\theta_y$ and $\theta_z$. This phase twisting
plays the role of the wavenumbers $k_x$, $k_y$ and $k_z$.
In this case, as the disorder generally breaks the ${\cal I}$-symmetry, 
the disorder brings about a gapless phase. 
This is an interesting question, and is beyond the scope of the
present paper.

\begin{acknowledgments}
We are grateful to R.~Shindou, L.~Balents, and X.-L. Qi 
for fruitful discussions.
This research is supported in part 
by Grant-in-Aids  
from the Ministry of Education,
Culture, Sports, Science and Technology of Japan.  

\end{acknowledgments}

\appendix
\section{General description of the gap-closing points in 
$m$-$\mathbf{k}$ space}
In this Appendix we assume that the bands are nondegenerate almost 
everywhere in $\mathbf{k}$ space. It allows existence of
isolated $\mathbf{k}$ points with band degeneracy;
meanwhile, the degeneracy in an extended region in $\mathbf{k}$ space,
such as Kramers degeneracy in systems with ${\cal I}$- and time-reversal symmetry 
is excluded. 
The vectors $\mathbf{A}_{\alpha}(\mathbf{k})$, $\mathbf{B}_{\alpha}
(\mathbf{k})$
are defined in the three-dimensional $\mathbf{k}$ space.
To study behaviors of the gap-closing point by changing $m$,
it is convenient to consider a four-dimensional $(m,\mathbf{k})$ space.
Let us write
\begin{equation}
k_{0}\equiv m,
\end{equation}
and we define the following 4-vectors
\begin{align}
&A_{\alpha, i}(k)\equiv A_{\alpha,i}
(m,\mathbf{k})=-i\langle \psi_{\alpha}(m,\mathbf{k})|\frac{\partial}{\partial
k_{i}}|
\psi_{\alpha}(m,\mathbf{k})\rangle,\\
&B_{\alpha,ij}(k)\equiv B_{\alpha,ij}
(m,\mathbf{k})=\frac{\partial}{\partial
k_{i}}A_{\alpha,j}(m,\mathbf{k})-\frac{\partial}{\partial
k_{j}}A_{\alpha,i}(m,\mathbf{k}),
\end{align}
where $k=(k_0,k_1,k_2,k_3)=(m,\mathbf{k})$ and $i=0,1,2,3$.
This corresponds to the vectors in Eqs.~(\ref{eq:A})(\ref{eq:B}) 
as $\mathbf{A}_{\alpha}
(\mathbf{k})=(A_1,A_2,A_3)$,
$\mathbf{B}_{\alpha}(\mathbf{k})=(B_{23},B_{31},B_{12})$.
We omit the band index $\alpha$  henceforth, unless necessary. 
The monopole density $\rho(\mathbf{k})$ becomes a 4-vector $\rho=(\rho_0,
\rho_1,\rho_2,\rho_3)$,
where
\begin{equation}
\rho_{l}=\frac{1}{2\pi}\epsilon_{lijk}\frac{\partial}{\partial k_{i}}
\frac{\partial}{\partial k_{j}}A_{k}.
\label{eq:rhol}\end{equation}
and $\epsilon_{lijk}$ is the totally antisymmetric tensor with 
$\epsilon_{0123}=1$. We note that $\rho(\mathbf{k})=\rho_{0}$.
From Eq.~(\ref{eq:rhol}), if the band $\alpha$ is not degenerate with 
other bands
at $(m,\mathbf{k})$, $A_{i}(k)$ is analytic, and 
$\rho_{l}(m,\mathbf{k})$ identically vanishes. 
This does not apply when 
the band is degenerate with other bands at some $\mathbf{k}$ points; 
at such degenerate points,
$\rho_l$ has a $\delta$-function singularity, as we see below. 
Such points form a curve in $(m,\mathbf{k})$ space because 
the gapless condition consists of three equations, i.e. the codimension 
is three \cite{vonNeumann29,Herring37} in this case. We call this gapless curve as a ``string''. 
We will see that the 4-vector $\rho$ describes a current inside the string,
and the total ``current'' inside the string is an integer. 
In general this ``current'' is unity.

To see how the singularity
appears, we show the following; (a) a surface integral of $\rho_{l}$ over
a closed three-dimensional hypersurface $V$ in $(m,\mathbf{k})$ space is zero
(i.e. $\rho$ is divergence-free), and 
(b) a surface integral of $\rho_{l}$ over
an open 3-dimensional hypersurface $\tilde{V}$
in $(m,\mathbf{k})$ space is quantized. 
To see (a) we use the Gauss theorem.
\begin{equation}
\int_{V}d\sigma^{ijk}\epsilon_{lijk}\rho_{l}=
\frac{1}{2\pi}
\int_{\partial V}d\sigma^{jk}
\frac{\partial}{\partial k_{j}}A_{k}=0,
\end{equation}
because $\partial V$ is null. This proof relies on the fact 
that
$B_{jk}=
\frac{\partial}{\partial k_{j}}A_{k}-\frac{\partial}{\partial k_{k}}A_{j}$ 
is gauge invariant. Next we show (b).
\begin{equation}
\frac{1}{3!}
\int_{\tilde{V}}d\sigma^{ijk}\epsilon_{lijk}\rho_{l}=
\frac{1}{2\pi}
\int_{\partial \tilde{V}}d\sigma^{jk}
\frac{\partial}{\partial k_{j}}A_{k}
\end{equation}
This quantity is an integer, representing a monopole 
charge inside $\partial \tilde{V}$. 
To show this we use the Stokes theorem to this expression. 
In general, when the closed two-dimensional surface $\partial \tilde{V}$ 
encloses a degeneracy point, the field $A_{k}$ cannot be expressed 
as a single function on the surface $\partial \tilde{V}$. Thus the Stokes
theorem can only be applied after dividing the surface $\partial \tilde{V}$ 
into
pieces, on each of which the wavefunction (and the field $A_i$) is
continuous\cite{Kohmoto85}. 
The resulting formula is expressed in terms of a phase
difference 
between the neighboring pieces. Because the phase of the wavefunctions
allows a difference of a multiple of $2\pi$, this quantity becomes an integer.
Thus from (a)(b), we have shown that $\rho_l$ describes a (divergence-free) 
``current'' 
in the string, with 
its current being quantized inside the string.
This kind of discussion with  
patching of the wavefunction
is discussed in the context of the Berry phase
\cite{Berry84,Volovik}, and 
also in the context of the quantum Hall effect
\cite{Kohmoto85}, and magnetic superconductor \cite{Murakami03b}.

\begin{figure}[htb]
\includegraphics[scale=0.5]{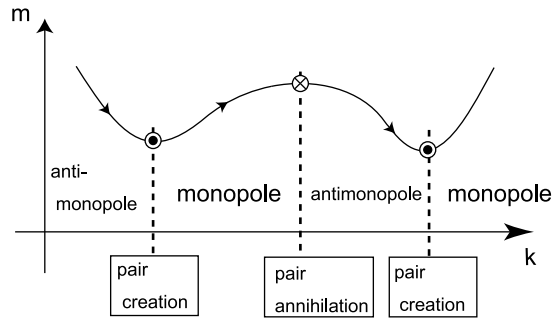}
\caption{The string which is a set of gapless points in the $(m,\mathbf{k})$ 
space. The arrow describes the direction of the flow vector $\rho$.}
\label{fig:string}
\end{figure}

The expression of $\rho_l$ is therefore given by  
\begin{equation}
\rho_{l}=
\int ds \frac{dK_{l}(s)}{ds}\delta^{(4)}(k-K(s))
\label{eq:rhol2}
\end{equation}
where $K(s)=(K_0(s),K_1(s),K_2(s),K_3(s))=(M(s),\mathbf{K}(s))$ 
describes a trajectory (i.e.\ string)
of the gap-closing point in the 4-dimensional space, $s$ is a parameter
along the string, 
and $\delta^{(4)}(k-K(s))=\prod_{i=0}^{3}\delta (k_{i}-K_{i}(s))$.
The divergence becomes zero because
\begin{align}
& \frac{\partial \rho_{l}}{\partial k_{l}}=
\int ds \frac{dK_{l}(s)}{ds}\frac{\partial}{\partial k_{l}}
\delta^{(4)}(k-K(s))\nonumber \\
&=
-\int ds \frac{dK_{l}(s)}{ds}\frac{\partial}{\partial K_{l}}
\delta^{(4)}(k-K(s))\nonumber \\
&=
-\int ds \frac{d}{ds}
\delta^{(4)}(k-K(s))=0.
\end{align}
The 0-th component of (\ref{eq:rhol2}) gives
\begin{align}
&\rho(m,\mathbf{k})=
\int ds \frac{dM(s)}{ds}\delta(m-M(s))
\delta^{(3)}(\mathbf{k}-\mathbf{K}(s))
\nonumber \\
&=
\sum_{s_{i}:m=M(s_{i})}
\mathrm{sgn}\left(\frac{dM(s)}{ds}\right)_{s=s_{i}}
\delta^{(3)}(\mathbf{k}-\mathbf{K}(s_{i})),
\end{align}
where the summation is taken over $s=s_i$ which satisfies $m=M(s_i)$ for given 
$m$.
By equating this with
\begin{equation}
\rho(m,\mathbf{k})=\sum_{s_{i}:m=M(s_{i})}
q_{i}
\delta^{(3)}(\mathbf{k}-\mathbf{K}(s_{i})),
\end{equation}
we get the monopole charge 
to be $q_{i}=\mathrm{sgn}\left(\frac{dM(s)}{ds}\right)_{s=s_{i}}$.
This means that when the ``current'' in the string is going in the 
increasing direction of $m$, it appears as a monopole ($q=1$), whereas
the decreasing direction of $m$ corresponds to an antimonopole ($q=-1$),
as shown in Fig.~\ref{fig:string}

We now return to the present case of the transition between 
the QSH and the insulator phases.
Recall that the string is confined in a restricted region $m_{1}\leq
m\leq m_2$ in the $m$ direction. Therefore the ``string'' becomes
a ``loop''. Then it follows that
at $m=m_{1}$ the ``string'' changes its direction in the $m$ space, 
which appears as a pair creation of monopole and antimonopole.
Similarly at $m=m_{2}$ a pair annihilation of monopole and 
antimonopole results. 

In the particular cases with time-reversal invariance, as we are 
interested in, the pair creation and pair annihilation appears symmetrically
with respect to $\mathbf{k}=\bm{\Gamma}_{i}$. 
Therefore, we expect that the number of loops should be one in 
the simplest case will be as shown in  Fig.~\ref{fig:monopole}, 
and it is indeed realized in the Fu-Kane-Mele model with the 
$\lambda_v$ term.

\section{Calculation of the gapless phase in
the Fu-Kane-Mele model with staggered on-site potential}
The Hamiltonian matrix for the 
the Fu-Kane-Mele model with staggered on-site potential $\lambda_v$
is written as
\begin{equation}
H(\mathbf{k})=\left(\begin{array}{cc}
\lambda_v\mathbf{1}
+\sum_{i=1}^{3}F_{i}\sigma_{i} & F_{0}\mathbf{1}\\
F_{0}^{*}\mathbf{1} & -\lambda_v\mathbf{1}
-\sum_{i=1}^{3}F_{i}\sigma_{i} 
\end{array}
\right)
\end{equation}
where $\sigma_{i}$ are the Pauli matrices, and 
$\mathbf{1}$ is the 2$\times$ 2
identity matrix. The coefficients are given by
\begin{align}
&F_{0}=t_{1}e^{ia(k_y +k_z)/2}+t_{2}e^{ia(k_z +k_x)/2}+
t_{3}e^{ia(k_x +k_y)/2}+t_{4},
\\
&F_{x}=-4\lambda_{\mathrm{SO}}\sin\frac{k_{x}a}{2}
\left(\cos\frac{k_{z}a}{2}-\cos\frac{k_{y}a}{2}\right),
\end{align}
and $F_{y}$, $F_{z}$ are given similarly to $F_{x}$ after cyclic
permutation of the subscripts $x$, $y$ and $z$. 
We write $\mathbf{F}=(F_x, F_y, F_z)$ for brevity.
We study when and how the gap closes as the parameters change.
Because the gap is between the second and the third bands, we need
a condition when the second and third bands have identical 
eigenenergies. As the codimension is three in this case, 
the condition should be expressed as three equations, determining
a string in $(m,\mathbf{k})$ space. 
First we note that 
the spectrum of
$H$ is symmetric with respect to $E=0$, as follows from 
\begin{equation}
UH(\mathbf{k})U^{-1}=-^{t}H(\mathbf{k}),
\end{equation} 
where the superscript ${}^{t}$ denotes matrix transposition, and 
the unitary 
matrix $U$ is given by 
\begin{equation}
U=\sigma_y\otimes \mathbf{1}=
\left(
\begin{array}{cccc}
&&-i&\\&&&-i\\i&&&\\&i&&
\end{array}
\right).
\end{equation}
Thus the gap closes if and only if one of the eigenvalues of 
$H$ vanishes, namely, $\mathrm{Det}H(\mathbf{k})=0$. 
This renders to
\begin{equation}
|F_{0}|^{4}+2|F_{0}|^{2}(\lambda_{v}^{2}+|\mathbf{F}|^{2})
+(\lambda_{v}^{2}-|\mathbf{F}|^{2})^{2}=0,
\end{equation}
namely,
\begin{equation}
\mathrm{Re}F_{0}=0,\ \mathrm{Im}F_{0}=0,\ 
 \lambda_{v}^{2}=|\mathbf{F}|^{2}.
\label{eq:gapless}
\end{equation}
For given model parameters $t_{i}$, $\lambda_{v}$, 
the coupled equations (\ref{eq:gapless}) gives
gapless points in $\mathbf{k}$-space, if any.
Such gapless points are stable against small changes of 
parameters, as we can see as follows. 
For brevity, let us write the three conditions in Eq.~(\ref{eq:gapless}) as
$g_i(\mathbf{k},\mathbf{t})=0$ ($i=1,2,3)$, where $\mathbf{t}$ denotes the
set of model parameters.
Let $\mathbf{k}_{0}$ be the wavenumber of one of the gapless points. 
When the model parameters are changed, the values of $g_i$ change 
accordingly: 
$g_i\rightarrow g_i+\delta g_i$. The gapless points are 
then expected to move ($\mathbf{k}=\mathbf{k}_{0}\rightarrow\mathbf{k}=
\mathbf{k}_{0}+\delta \mathbf{k}_{0}$). Because
the gapless conditions are expanded as
\begin{equation}
\nabla_{\mathbf{k}}g_i(\mathbf{k}_{0})
\cdot \delta \mathbf{k}_{0}+\delta g_i(\mathbf{k}_{0})=0,
\end{equation}
there always exists $\delta\mathbf{k}_{0}$ satisfying these three coupled
linear equations. Thus a perturbative change in  
system parameters moves the gapless points in $\mathbf{k}$-space,
without removing them.

In particular, when $\lambda_{v}=0$, the system is ${\cal I}$-symmetric.
The third condition in Eqs.~(\ref{eq:gapless}) then gives the wavenumber 
$\mathbf{k}$ to be one of the $X^{r}$ points;  the other conditions lead
us to the phase diagram in Ref.~\onlinecite{Fu06b} (also in 
Fig.~\ref{fig:phase-dia1}(a)) easily.

\end{document}